\documentclass[aps,prb,superscriptaddress,twocolumn]{revtex4}
\usepackage{graphicx}
\usepackage{times}
\usepackage{tabularx}
\usepackage{amsmath}
\usepackage{amsthm}
\usepackage{amssymb}
\usepackage{amsbsy}
\newcommand{\abs}[1]{\vert #1 \vert}
\newcommand{\bra}[1]{\langle #1 \vert}
\newcommand{\ket}[1]{\vert #1 \rangle}

\newcommand{\sgn}{\text{sgn}}

\begin{document}
\title{Probability and (Braiding) Statistics in Majorana Nanowires}
\author{David J. Clarke}
\affiliation{Condensed Matter Theory Center, Department of Physics, University of Maryland, College Park}
\affiliation{Joint Quantum Institute, University of Maryland, College Park}
\affiliation{Station Q Maryland}
\author{Jay D. Sau}
\affiliation{Condensed Matter Theory Center, Department of Physics, University of Maryland, College Park}
\affiliation{Joint Quantum Institute, University of Maryland, College Park}
\affiliation{Station Q Maryland}
\author{S. Das Sarma}
\affiliation{Condensed Matter Theory Center, Department of Physics, University of Maryland, College Park}
\affiliation{Joint Quantum Institute, University of Maryland, College Park}
\affiliation{Station Q Maryland}

\begin{abstract}
    Given recent progress in the realization of Majorana zero modes in semiconducting nanowires with proximity-induced superconductivity, a crucial next step is to attempt an experimental demonstration of the predicted braiding statistics associated with the Majorana mode. Such a demonstration should, in principle, confirm that the experimentally observed zero-bias anomalies are indeed due to the presence of anyonic Majorana zero modes. Moreover, such a demonstration would be a breakthrough at the level of fundamental physics: the first clear demonstration of a non-Abelian excitation. It is therefore important to clarify the expected signals of Majorana physics in the braiding context, and to differentiate these signals from those that might also arise in non-topological variants of the same system. A definitive and critical distinction between signals expected in topological (i.e. anyonic) and non-topological (i.e. trivial) situations is therefore essential for future progress in the field. In this manuscript, we carefully examine the expected signals of proposed braiding and fusion experiments in topological and non-topological variants of the experimental nanowire systems in which Majoranas are predicted to occur. We point out situations where `trivial' and `anyonic' signatures may be qualitatively similar experimentally, necessitating a certain level of caution in the interpretation of various proposed fusion and braiding experiments. We find in particular that braiding experiments consisting of full braids (two Majorana exchanges) are better at distinguishing between topological and non-topological systems than fusion experiments or experiments with an odd number of Majorana exchanges. Successful fusion experiments, particularly in nanowires where zero bias conductance peaks are also observed, can also provide strong evidence for the existence of Majorana modes, but such fusion evidence without a corresponding braiding success is not definitive.
\end{abstract}
\maketitle
\section{Introduction}
    Topological superconductors~\cite{Schnyder08} supporting Majorana zero modes~\cite{Read00, Kitaev01,Sengupta01} provide one of the simplest systems that are predicted to support non-Abelian statistics~\cite{Ivanov01}. Such non-Abelian statistics with the accompanying topological degeneracy associated with the Majorana zero modes (MZMs) may be used as the basis for topologically protected schemes for quantum computation~\cite{Nayak08,Alicea12a,Leijnse12,Beenakker2013a,Stanescu13,DasSarma15,Elliott15}. The theoretically proposed~\cite{Sau10a,Lutchyn10,Oreg10}  semiconductor-based structures for realizing topological superconductors have led to encouraging experimental results~\cite{Mourik12,Das12,Deng12,Rokhinson12,Finck12,Churchill13,Chang15,Zhang16,Albrecht16}, suggesting that such semiconductor nanowire devices might be a viable path to eventual fault-tolerant topological quantum computation~\cite{Nayak08,Alicea12a,Leijnse12,Beenakker2013a,Stanescu13,DasSarma15,Elliott15,Alicea11,Sau10c,Clarke11b,vanHeck12,Sau11b,Hyart13,Clarke16,Plugge16,Karzig16}.

    The optimistic experimental results involving the observation of the predicted zero bias conductance peaks in nanowire tunneling transport measurements have encouraged further theoretical proposals to demonstrate ideas related to testing braiding and non-Abelian statistics~\cite{Alicea11,Sau11b,Hyart13,Clarke16}. In particular, proposals to directly test the non-Abelian fusion rules associated with Majorana zero modes~\cite{Ruhman14,Aasen16} substantially simplify the necessary device design relative to that required for braiding experiments, encouraging experimental groups to undertake a search for this simplest non-trivial non-Abelian signature. However, despite the initial and repeated success in observing the predicted zero bias conductance peaks (ZBCPs) associated with the existence of Majorana zero modes~\cite{Mourik12,Das12,Deng12,Finck12,Churchill13,Chang15,Zhang16,Albrecht16}, the expected precise and robust quantization of the conductance at zero bias,\cite{Sengupta01,Nayak08,Alicea12a,Leijnse12,Beenakker2013a,Stanescu13,DasSarma15,Elliott15} which is one of the definitive characteristics of Majorana zero modes, remains elusive even after five years of substantial experimental effort\cite{Das12,Deng12,Rokhinson12,Finck12,Churchill13,Chang15,Zhang16,Albrecht16} following the initial observation of a zero-bias peak.\cite{Mourik12} This raises the possibility that the observed zero bias conductance may arise from physics other than non-Abelian Majorana zero modes such as disorder induced zero energy states, weak antilocalization, multiple unsplit Majorana zero modes~\cite{Bagrets12,liu2012zero,kells2012near,pikulin2012zero} or other unknown reasons unrelated to Majorana physics.  This is particulary worrisome since the zero bias conductance peak is a necessary but by no means sufficient condition for the existence of non-Abelian Majorana modes. On the other hand, it is not evident that any of these non-Majorana possibilities are quantitatively consistent with the experimentally observed signatures of Majorana zero modes. Clearly a detailed understanding of the zero bias conductance is still incomplete\cite{DasSarma16,Liu16} despite the essential simplicity of the tunneling conductance measurement compared to the substantially more complex proposals involved in fusion and braiding~\cite{Alicea11,Sau11b,Hyart13,Clarke16,Ruhman14,Aasen16}.  This is true despite the essential simplicity of the conductance experiment, which has permitted a detailed theoretical analysis
    of the conductance in systems with disorder, interaction, dissipation and most importantly even non-topological systems with no Majorana modes~\cite{Fidkowski12,adagideli2014effects,DasSarma16,liu2012zero,LinC12, Moore16, LeeEJH14}. An important possible scenario is that the existing tunneling transport measurements indeed observe nanowire Majorana zero modes, but that realistic effects in laboratory systems act to allow significant couplings to and between the Majorana modes that would not be present in an ideal topological system.\cite{Moore16} Such modes may be described as quasi- or almost-MZMs, and the important question then becomes whether such quasi-MZMs carry non-Abelian statistics or not as manifested in braiding experiments.

    Thus, while it is clear that ideal Majorana zero modes should have interesting features in braiding and fusion experiments~\cite{Alicea11,Sau11b,Hyart13,Clarke16,Ruhman14,Aasen16}, the complex nature of the various proposed experiments requires a deeper analysis and a broader understanding in light of the zero-bias conductance observations that do not report the theoretically predicted quantized peak. Specifically, it is crucial to understand in detail the results of these proposed fusion and braiding experiments for more generic and realistic systems as compared to ideal topological ones. For example, one could ask whether systems that possess low energy fermionic Andreev bound states as opposed to Majorana zero modes respond qualitatively differently to the proposed fusion and braiding experiments. It has already been claimed that such low-lying accidental Andreev bound states may give rise to zero bias conductance peaks similar to the Majorana peaks\cite{LeeEJH14}, making this question a key experimentally relevant issue. The same issue is also germane if the zero bias peak arises from an almost-Majorana mode comprising an overlap of two (or more) Majorana modes localized at different spatial regions of the nanowire (not necessarily at the endpoints). One of the central goals of the current paper is to provide an extensive characterization of the fusion and braiding experiment so as to be able to answer such questions, in the process providing clear guidelines about distinguishing non-Abelian topological and trivial non-topological features in proposed future fusion or braiding experiments on nanowires.

    This paper is organized as follows: In Sec.~\ref{Sec: Fusion}, we discuss recent proposals\cite{Ruhman14, Aasen16} for measuring the fusion rules of MZM defects. We find that the key signature of MZMs in these experiments--a 50\% probability of measuring an odd parity--is generically reproduced in a random system (independent of its intrinsic topological properties), and is even more likely in a system that is already known to display zero bias peaks (independent of whether the peak arises from non-Abelian MZMs, quasi-MZMs or accidental Andreev states). This particular way of characterizing non-Abelian Majorana modes through simple fusion experiments is therefore problematic since a trivial non-topological system might manifest the same behavior. We draw an analogy to a spin-1/2 particle precessing in a time-varying magnetic field, and use this analogy to highlight the special characteristic of the `true' (\emph{i.e} non-Abelian) Majorana system. In Sec.~\ref{Sec: Braiding}, we move on to braiding experiments, outlining a set of assumptions that allow us to narrow down the possible results of carrying out nominal braid operations through adiabatic evolution in non-topological systems. We find a remarkable coincidence in the braiding result for a system with one topological wire and one wire having `accidental' degeneracy. Such a system reproduces exactly the 50\% probability of measuring odd or even parity after a single braid, despite the presence of non-topological couplings. Again, this is a possibility one must keep in mind in interpreting future experiments searching for purely topological braiding effects associated with non-Abelian anyonic excitations. In Sec.~\ref{Sec: Poisoning}, we deal with the implications of quasiparticle poisoning on the degradation of the braiding signal. We conclude in Sec.~\ref{Sec: Outlook} with a discussion of the outlook for ongoing Majorana braiding experiments, with emphasis on the caution and care necessary in the interpretation of proposed fusion/braiding experiments.
\section{Fusion experiments}\label{Sec: Fusion}

    We begin with an analysis of the simple fusion experiment described conceptually in Ruhman et al.\cite{Ruhman14} and further extended and elaborated by Aasen et al.\cite{Aasen16}  In this experiment, two regions of a superconducting one dimensional system, each ostensibly containing Majorana zero modes at their endpoints, are placed end to end. Preparatory measurements are made to assure that the total fermion parity of the combined system is even. A strong tunnel coupling between the ends of the two subsystems prepares a superposition of the fermion parity states $\ket{00}$ and $\ket{11}$, where each 0 or 1 represents the fermion parity of the left or right regions, respectively. If the system actually contains Majorana zero modes localized at the endpoints of the subregions, the superposition is expected to be equal, so that when the system is broken apart\footnote{We refer to this `breaking apart' as `cutting' in our discussion and, therefore, the 'cutting rate' denoting the rate at which this `breaking apart' occurs is an important time scale or energy scale in the fusion problem.} by removing the coupling on a time scale rapid compared to any remaining Majorana splitting, the state $\ket{00}$ will be measured 50\% of the time, and the state $\ket{11}$ will be measured 50\% of the time. This has been argued to represent a distinct consequence of the fusion rules for Majorana zero modes,\cite{Aasen16, Ruhman14} and therefore as indirect evidence supporting the non-Abelian nature of these excitations. The key question to be addressed here is whether such a fusion measurement intrinsically conveys any \emph{more} information than the already observed zero bias conductance peak arising from low energy (topological or non-topological) subgap modes.\cite{Mourik12,Das12,Deng12,Rokhinson12,Finck12,Churchill13,Chang15,Zhang16,Albrecht16,LeeEJH14}

    Here we take a somewhat egalitarian approach to our analysis of this experimental proposal. First, we note that the zero bias peaks observed in nanowires built with the intention of hosting Majorana zero modes\cite{Mourik12} represent, at the very least, direct evidence for the presence of low-lying energy states at the endpoints of these wires. In the simplest form of our analysis, we assume that each subregion has a single low-lying fermionic mode, with associated energy scales $\epsilon_1$ and $\epsilon_2$ respectively. Fermion tunneling between the two regions couples states with the same overall fermion parity. Finally, a (likely small) cross-capacitance between the two regions alters the energy by $\epsilon_C$ when both fermionic modes are occupied. Our Hamiltonian is therefore
    \begin{equation}\label{eq:simple}
        H=\left(\begin{array}{c} \ket{00}\\\ket{01}\\\ket{10}\\\ket{11}\end{array}\right)^T
            \left(\begin{array}{cccc}
                0 & 0 & 0 & h_1\\
                0 & \epsilon_2 & h_2 & 0\\
                0 & h_2^*  & \epsilon_1 & 0\\
                h_1^* & 0 & 0 & \epsilon\\
            \end{array}\right)\left(\begin{array}{c} \bra{00}\\\bra{01}\\\bra{10}\\\bra{11}\end{array}\right)
    \end{equation}
    where $\epsilon=\epsilon_1+\epsilon_2+\epsilon_C$. Here $h_{1,2}$ are the appropriate tunnel couplings between the two regions.

    Due to fermion parity conservation, we may separate this Hamiltonian into even and odd parity blocks, from which we may easily determine the eigenstates

    \begin{eqnarray}
        \ket{v_1}=\cos{\alpha}\ket{00}+e^{i\phi_1}\sin{\alpha}\ket{11}\nonumber\\
        \ket{v_2}=\sin{\alpha}\ket{00}-e^{i\phi_1}\cos{\alpha}\ket{11}\nonumber\\
        \ket{v_3}=\cos{\beta}\ket{01}+e^{i\phi_2}\sin{\beta}\ket{10}\nonumber\\
        \ket{v_4}=\sin{\beta}\ket{01}-e^{i\phi_2}\cos{\beta}\ket{10}\nonumber\\
    \end{eqnarray}
    where $\tan{2\alpha}=\frac{2\abs{h_1}}{\epsilon}$, $\tan{2\beta}=\frac{2\abs{h_2}}{\epsilon_2-\epsilon_1}$, and \mbox{$e^{2i\phi_{1,2}}=h_{1,2}/h_{1,2}^*$}.

    In the fusion proposal of Ref.~\onlinecite{Aasen16}, the two subregions are separated on a timescale that is adiabatic with respect to all but the lowest energy mode, which would be the zero-energy MZM in the ideal topological scenario. With respect to this last mode, the experiment must proceed non-adiabatically (\emph{i.e.} suddenly) in order to achieve a result that is distinct from the control experiment.\cite{Hell16} While in the most general system it may be difficult to independently tune the parameters of the above Hamiltonian, the zero-bias peaks observed in transport experiments have proven relatively insensitive to the voltage present on a back-gate near the end of the proximitized wire.\cite{Mourik12} We therefore assume here that the voltage change necessary to separate the regions by depleting an intermediate tunnel barrier does not significantly change the splitting of modes within each subregion. If this is untrue, then one must know the detailed voltage dependence of the ZBCP to make further progress in this model, but the relative insensitivity of the zero bias peak to local perturbations in voltage is a necessary feature of the topological system\cite{Nayak08,Alicea12a,Leijnse12,Beenakker2013a,Stanescu13,DasSarma15,Elliott15} that may be (and has been\cite{Mourik12}) tested in transport experiments.

    If the tunnel coupling were to be abruptly turned off in the above Hamiltonian, the state $\ket{00}$ would be measured (\emph{e.g.} with a charge sensing measurement as outlined by Ref.~\onlinecite{Aasen16}) with probability
    \begin{equation}\label{eq:onemode}
        \mathcal{P}_{00}=\frac{1}{2}\left(1\pm \cos{2\alpha}\right)=\frac{1}{2}\left(1\pm \frac{\epsilon}{\sqrt{4\abs{h_1}^2+\epsilon^2}}\right)
    \end{equation}
    For systems with true Majorana zero modes, we expect $\epsilon_1=\epsilon_2=0$, and $\epsilon_C\ll h_{1,2}$.  However, in this abrupt approximation, the condition $\epsilon \ll h_1 \ll 1/\tau$, where $\tau$ is the time taken to make the cut, will lead to a measured probability $\mathcal{P}_{00}\sim 1/2$.\footnote{If $h_1\gtrsim 1/\tau$, then the sudden approximation fails for part of the evolution. In this case $h_1$ is roughly replaced by $1/\tau$ in Eq.~\ref{eq:onemode}.} More generally, in the presence of multiple (perhaps low energy) fermionic modes on each side of the tunnel junction (and in the sudden approximation), a probability near 1/2 will be measured whenever the tunnel coupling scale is much larger than the energy of the lowest lying excited fermionic state on either side of the junction, and much smaller than the energy scale set by the cutoff rate or the next lowest fermionic mode. Thus, a trivial situation with multiple low-energy fermions will mimic the Majorana situation in such a measurement. The important point to emphasize here is that the experiment manifesting the observation of 1/2 probability by itself has no way of assuring the absence of such non-topological low-energy fermionic modes that would lead to this observation. What it can establish is the existence of modes which have lower energies (and are thus diabatic) with respect to the cutting rate. The fusion experiment itself has no direct way of establishing the topological nature of these low energy modes.

    \begin{figure}
    \centering
    \includegraphics[trim=4cm 8cm 4cm 8cm, clip,width=.5\columnwidth]{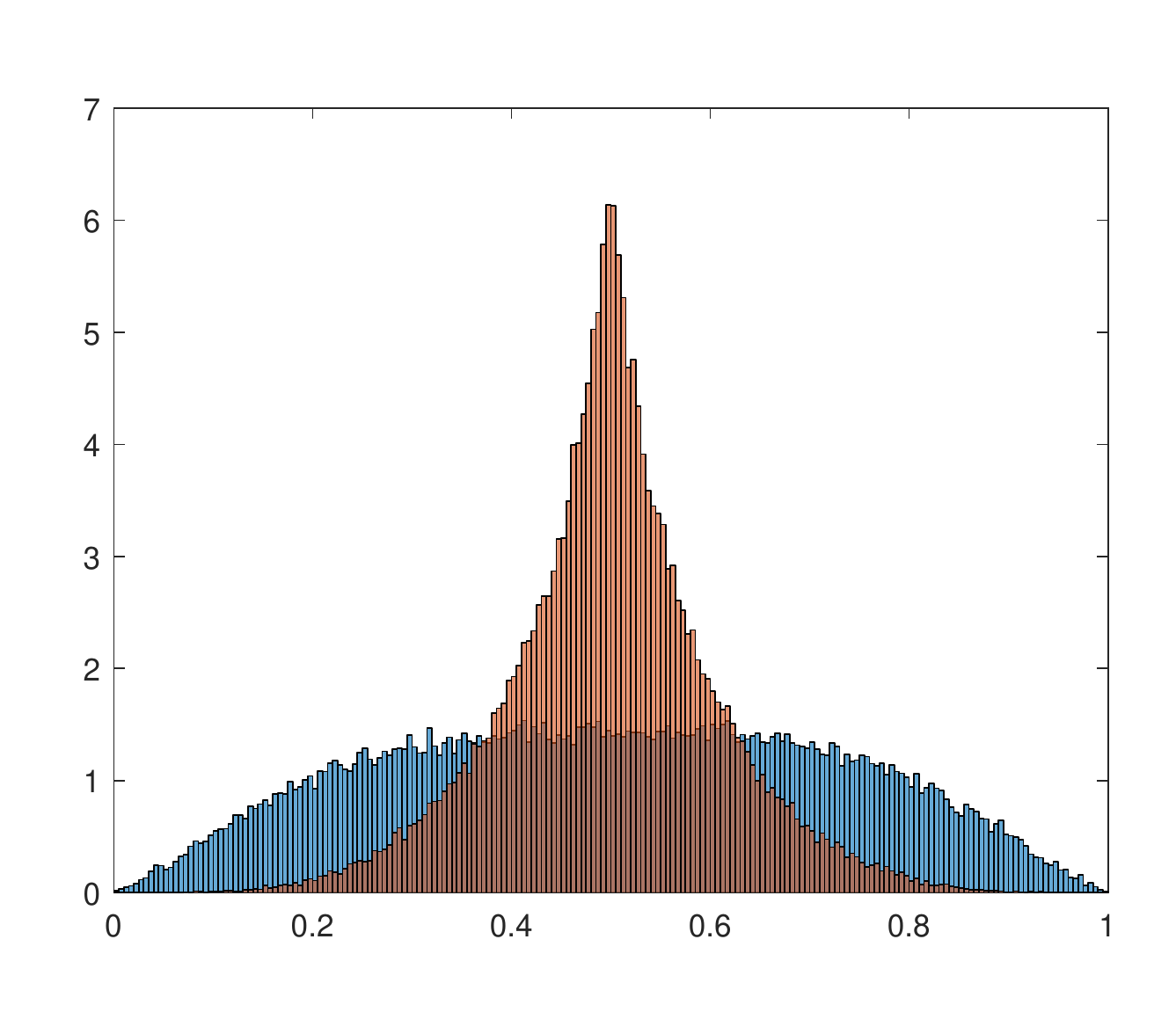}\includegraphics[trim=4cm 8cm 4cm 8cm, clip,width=.5\columnwidth]{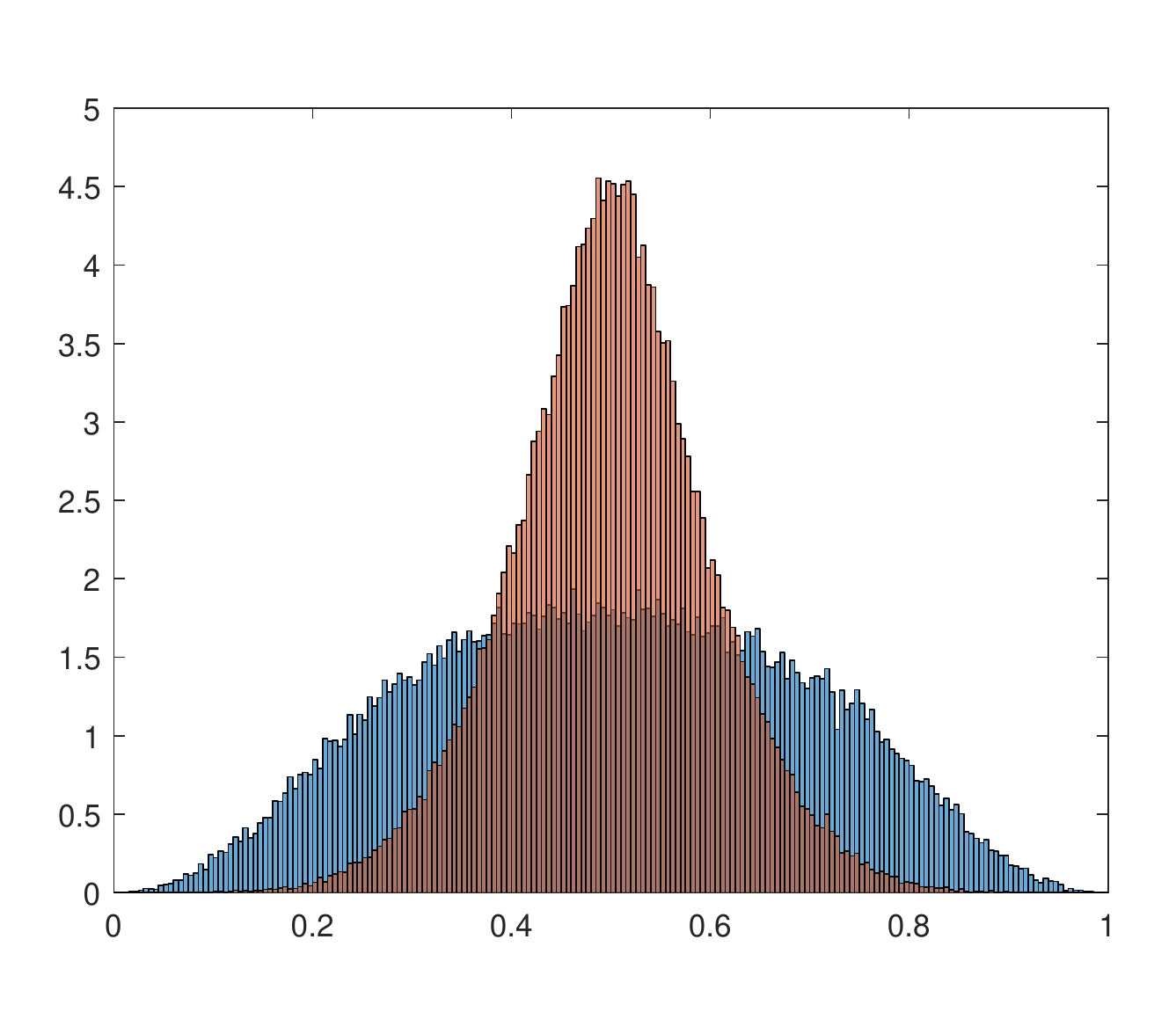}
    \includegraphics[trim=4cm 8cm 4cm 8cm, clip,width=.5\columnwidth]{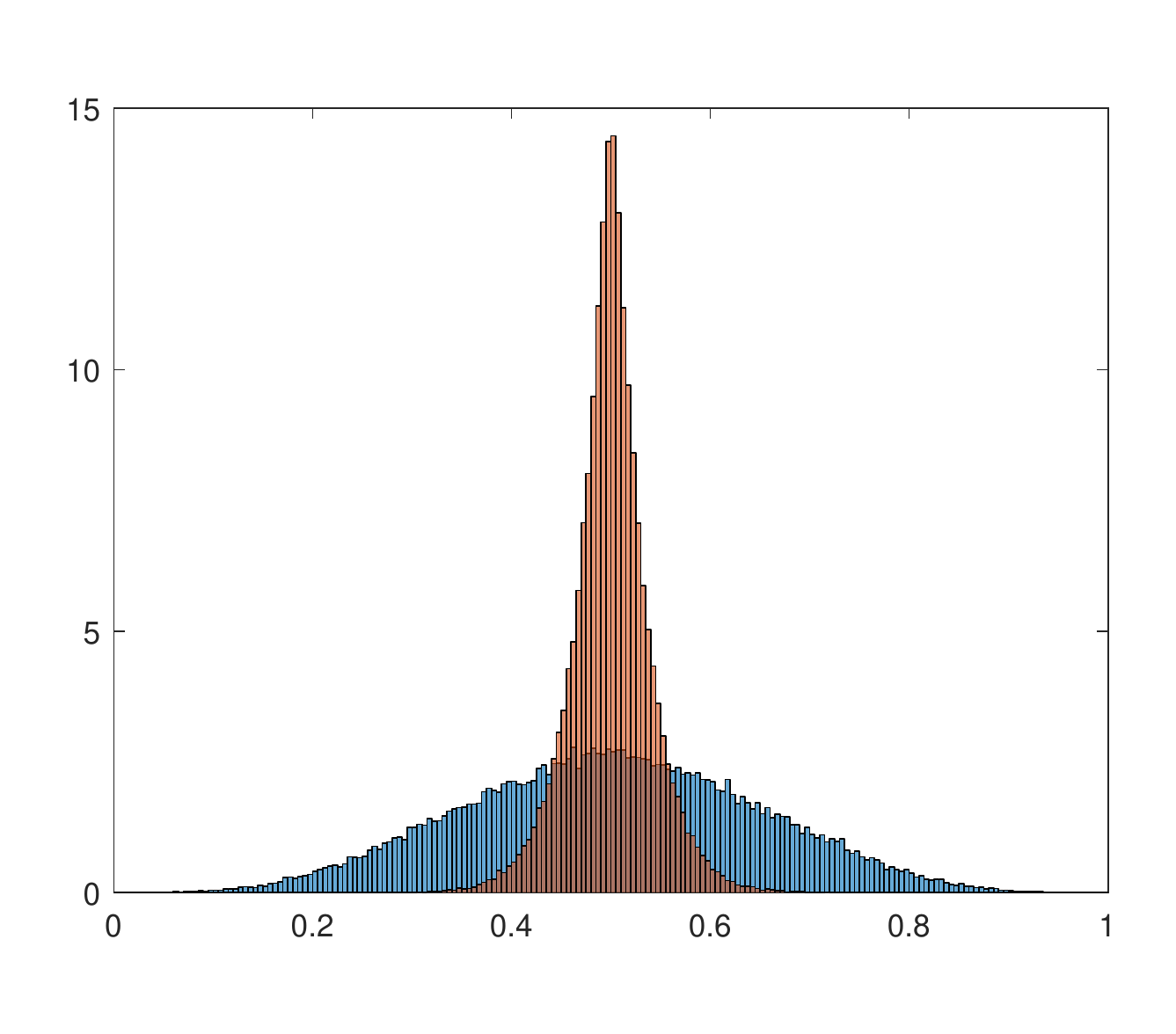}\includegraphics[trim=4cm 8cm 4cm 8cm, clip,width=.5\columnwidth]{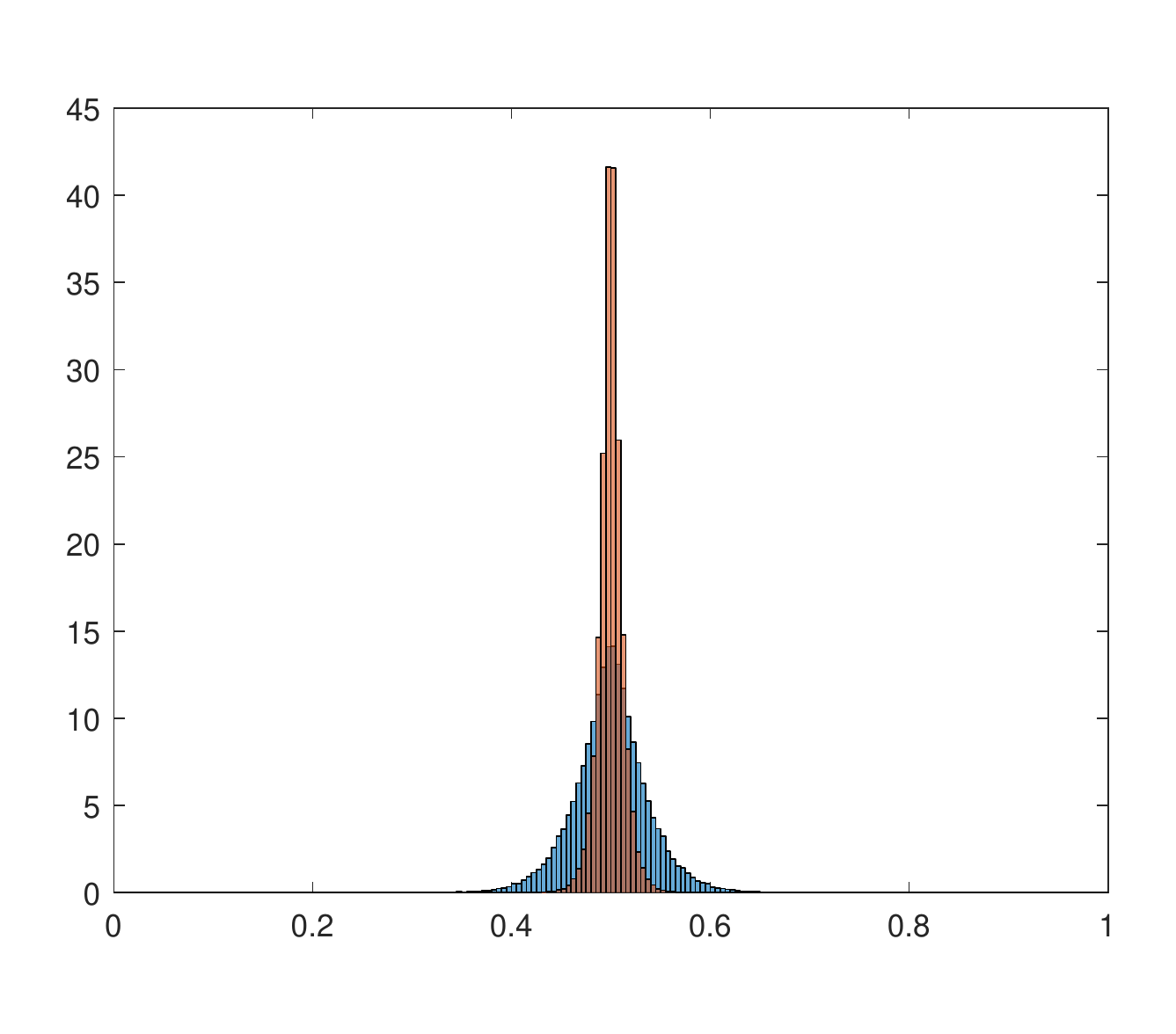}
        \caption{(Color Online) Histograms of the probability of measuring even parity in a fusion experiment in which several modes (localized to subregions $A$ and $B$) have energy well below the frequency with which the two regions are cut apart. The histograms show a peak around a probability of 50\%. The histograms in blue are for gaussian distributed random Hamiltonians. Those in red, showing a narrower peak, result when the couplings are chosen to create an exact zero energy mode within each subregion.  (This mode is generically of non-topological origin within our simulation, as couplings are allowed between regions that would split the energy away from zero.)  (Top Row, Left) $N=3$ low energy fermions split into $n_A=2$, $n_B=1$ (Top Row, Right) $N=4$ low energy fermions split into $n_A=3$, $n_B=1$. (Bottom Row, Left) $N=4$ low energy fermions split into $n_A=2$, $n_B=2$. (Bottom Row, Right) $N=8$ low energy fermions split into $n_A=4$, $n_B=4$.}  \label{Fig:Hist1}
    \end{figure}
    We may further explore the intuitive picture offered by the above model by explicitly including the possible additional low-lying modes mentioned above. We now allow several ($N$) low-lying fermionic modes in the system with random couplings, divided randomly into two subregions $A$ and $B$ with $n_A$ fermionic modes in region $A$ and $n_B=N-n_A$ modes in region $B$. The effective Hamiltonian describing the low-lying states is written as $H=i \sum_{ab}h_{ab}\gamma_{a}\gamma_b$, where $h_{ab}$ is an anti-symmetric real matrix and $\gamma_a$ represent a basis of Majorana operators describing the system. The single-particle density matrix of the system, $\rho_{ab}=i\langle\gamma_a\gamma_b\rangle$, is given by $\rho_{ab}=\sum_n \Psi_{na}^*\Psi_{n,b}$ where $\Psi_{n,a}$ are negative energy eigenstates of $h_{ab}$. Making a `sudden' approximation (as before) that the coupling between the two regions is turned off at a rate much higher than the energy scale of the coupling $h_{ab}$ of the low-energy space, the system develops conserved fermion parities $Q_A$ and $Q_B$ for each subregion. We again emphasize that this situation can occur even in systems that are definitively non-topological when low-energy modes (with energy scale below the cutting rate) are present. As we shall see, such systems are able to mimic some (but not all!) of the signals of braiding, as well, and care must be taken in ruling out the `false positives' that are likely to be present in realistic non-topological nanowires, especially if those wires are known to have zero-bias peaks in tunneling conductance.

    Assuming that such a multiple-mode situation arises from disorder and is therefore random, we compute the expectation $\langle Q_A \rangle$ over an ensemble of Gaussian random antisymmetric matrices. We restrict to the overall even-parity subsector, and plot the resulting histogram of $\mathcal{P}_{00}$ probabilities in blue in Fig.~\ref{Fig:Hist1}. Note the generic broad peak around $\mathcal{P}_{00}=1/2$ despite the fact that (by construction) there are generally no non-Abelian MZMs in the model leading to Fig.~\ref{Fig:Hist1} (and only low-energy fermionic excitations). The results shown in Fig. 1 explicitly demonstrate that the observation of a probability-1/2 outcome in the fusion measurement might be generic in the presence of many low energy modes (independent of their topological nature) arising from random disorder (which is always present in experiments) and cannot by itself be construed as definitive evidence for non-Abelian statistics.

    A further refinement of the model might include the information (known from transport data\cite{Mourik12,Das12,Deng12,Finck12,Churchill13,Chang15,Zhang16,Albrecht16}) that a zero-energy mode is independently present at the end of each wire region when the coupling between the two regions is turned off. We note that the existence of any fermionic mode $\hat{c}$ at exactly zero energy implies that there is an MZM $\gamma=\hat{c}+\hat{c}^\dagger$ (though this mode is not topologically derived in general). Therefore the refinement of the model may be accomplished by assuming that there is (at least) one Majorana operator in each sub-region that couples only to modes in the \emph{other} subregion.  We contrast this with the topological situation in which there exists at least one Majorana mode that does not couple to \emph{any} other modes, independent of the coupling between subregions.  Assuming this on-site condition remains even when the coupling turns on, the distribution of $\mathcal{P}_{00}$ becomes much more sharply peaked around $\mathcal{P}_{00}=1/2$ (red histograms of Fig.~\ref{Fig:Hist1}). There are two important caveats in this analysis. First, the transport experiments do not necessarily imply a mode at precisely zero energy-- indeed, the current experiments\cite{Mourik12,Das12,Deng12,Finck12,Churchill13,Chang15,Zhang16,Albrecht16} observe a ZBCP which is almost as broad as the topological gap itself implying the mode energy could be as large as half the superconducting gap. Second, the on-site couplings between Majorana modes may indeed drift as the inter-region coupling is turned on or off, counter to our initial assumption of independent control. We therefore expect the actual distribution of even-parity probabilities in a random multi-mode system that reproduces the transport results to interpolate between the red and blue histograms shown in Fig.~\ref{Fig:Hist1}.

    Thus far, our analysis has taken place in a `sudden' approximation: the energies associated with the low lying states are assumed to be small compared with the rate at which the coupling between the two regions goes to zero.  As the system is cut at a faster and faster rate, more modes become active and the probability distribution becomes more and more sharply peaked around $\mathcal{P}_{00}=1/2$. This should not be surprising, as at very high cutoff rates (which basically correspond to very high energies) we may expect the system to behave essentially as a fermi gas, with no preference for even or odd parity in the two subregions. At slower cutoff rates, the higher energy modes `freeze out,' leaving us with the few-mode effective model described above. If we cut slowly enough, even these last fermionic modes freeze out and leave the qubit polarized with a definite parity on each island at the end of the evolution. Such a slow cutting rate is thus important in the fusion experiment for establishing the Majorana zero mode, but, on the other hand, the rate cannot be so slow that the system decoheres (\emph{e.g.} through quasiparticle poisoning) or so slow that the residual coupling between Majorana modes can polarize the system (\emph{i.e} slower than the scale set by the Majorana splitting).\cite{Hell16,DasSarma12}  We note that tunneling transport on existing nanowires most often finds broad zero-bias conductance peaks  (sometimes of the order of the induced gap), which may indicate the generic presence of several low-energy fermionic modes in addition to one near zero energy. If this is indeed the source of the broadening, a $\mathcal{P}_{00}=1/2$ result may be expected unless the cut rate is slow enough to freeze out these modes.

    It is therefore clear that a $\mathcal{P}_{00}=1/2$ result in the fusion experiment is ultimately analogous to spectroscopy in the sense that it determines the proximity of the energy of low-lying states to zero energy. The measurement precision will depend on (and be roughly of the scale of) the experimental cut rate, although the precision cannot surpass that set by the quasiparticle poisoning rate. We may quantify the relevant time or energy scales by examining, e.g., the zero-bias peak data of Mourik et al. At a temperature of $60mK$, Mourik et al.\cite{Mourik12} measured a zero bias peak in their transport data with a full width at half maximum of $20\mu eV\sim 252 mK\gg 60mK$. This gives an approximate upper bound on the energy scale $\epsilon$ of the low energy mode or modes leading to the zero bias peak.  While the quality of the conductance data has improved in more recent experiments\cite{Zhang16}, the peak width remains of a similar order of magnitude.\cite{Liu16} One of the chief benefits of a fusion-type experiment is therefore the possibility of an improvement in the energy resolution of the low-lying states (i.e. by going to slower and slower cut rates). If an experiment using wires similar to those in transport experiments\cite{Mourik12,Das12,Deng12,Finck12,Churchill13,Chang15,Zhang16,Albrecht16} has a cut rate that is faster than the time scale set by the $\sim20\mu eV$ peak width (\emph{i.e.} taking a time shorter than $\sim0.016ns$ to perform the cut), a probability of $\mathcal{P}_{00}=1/2$ may be expected quite generically, as it is already known from those transport experiments that modes with energy $\lesssim 20\mu eV$ are present near the ends of the wire. Observing a peak in the fusion probability at 1/2 at a cut rate slower than the $20 \mu eV$ scale (taking longer than 0.016 ns) would confirm the existence of zero modes to a higher level of precision than these conductance experiments. At sufficiently slow cut rates the fusion experiment can be expected to lead to the adiabatic result. If a $\mathcal{P}_{00}=1/2$ is consistently seen, the cut rate would then determine a new upper bound on the splitting of the low energy states away from zero.

    Since $0.016 ns$ is indeed a rather short time scale, it is encouraging that successful fusion measurements may very well be able to decrease the uncertainty in the zero-mode considerably-- in fact (at zero temperature) a factor of $20$ improvement (i.e. the mode energy being constrained to within $1 \mu eV$ of zero) should be achievable by using experimental cut rates \mbox{$\sim$ 1ns}.  This will still not settle the question of the topological nature of these modes, but being able to determine with precision how low in energy the low energy modes really are will be a great improvement compared with the existing transport results where the bound on the mode energy from the peak width appears to be stuck near $~ 10 \mu eV$.

    Such an improved resolution may be able the resolve the Majorana splitting present in current nanowires. Following Aasen et al.,\cite{Aasen16} we may use the earlier theoretical work of Das Sarma et al.,\cite{DasSarma12} to put a conservative estimate of the Majorana splitting for a nanowire similar to those used in current experiments ($3\mu m$ long with a pairing energy of $1K$ and an induced coherence length of $500nm$) at $3mK\sim 0.25\mu eV$. This would suggest that in the case of ideal topological Majorana modes that follow the theory closely, the fusion experiment on these wires would cease to give values of $\mathcal{P}_{00}$ near $1/2$ for cutting time-scales longer than $\hbar/\epsilon\sim 1.5ns$. (This assumes that the Majorana modes are localized at the ends of the wire, which is by no means assured.\cite{Moore16}  Importantly, the fusion experiment does eliminate the normal metal lead present in transport experiments. It can therefore rule out sources of the zero-bias peak (such as the Kondo effect\cite{LeeEJH12}) that are not intrinsic to the nanowire but rather require the interaction of the lead with the nanowire. The utility of fusion experiments may increase in longer wires, which are expected to have a smaller splitting in the topological degeneracy (requiring a slower cutoff rate to resolve via fusion) but which may have similar resolution in transport. However, experimental data (even for tunneling transport) in such long nanowires does not yet exist (and the issue of disorder-induced low lying fermionic states may become more severe in longer wires).

    We emphasize again that the hard physical constraint to making the cut rate extremely slow (so that the process is adiabatic with respect to everything except for a strict zero energy mode) is that it must be faster than the quasiparticle poisoning rate, which can randomly change the parity. It is important to note that the control experiment described by Ref.~\onlinecite{Aasen16} can effectively determine whether quasiparticle poisoning is happening on the timescale of the experiment, thus avoiding a false positive. The best case scenario for the fusion experiment is therefore an energy resolution on the scale of the inverse quasiparticle poisoning time. This time may be extremely long\cite{vanWoerkom15,Higginbotham15}, but even in more conservative estimates should be long enough to resolve the above Majorana splitting.\cite{Rainis12} Whether the quasiparticle poisoning rate is indeed the functional resolution of the fusion experiment remains to be seen, and is beyond the scope of our analysis here.  We note, however, that we have not taken into account the effects of temperature, which may generically be expected to favor the maximally random result (\emph{i.e.} $P_{00}=1/2$ and which therefore can only decrease the experimental resolution. This is especially relevant when the expected splitting is far below the base temperature scale of the experiment, which has typically been $~50-60 mK$.\cite{Mourik12,Das12,Deng12,Finck12,Churchill13,Chang15,Zhang16,Albrecht16}

    Given our emphasis on the fact that fusion experiments do not uniquely determine a system to have topologically-derived MZMs, one may ask if there is anything special about a `true' Majorana system (at least with regard to this fusion experiment) beyond simply having a low lying fermionic mode such that $\epsilon=0$. We may illuminate the difference by returning to the simple model of Eq.~(\ref{eq:simple}), which may be mapped onto the Hamiltonian for a spin~$1/2$ evolving in a time-dependent magnetic field. The energy difference between even and odd occupation of the two subregions $\epsilon$ is mapped to the field $B_z$ in the $z$-direction, while $h_1=B_x+iB_y$. In the true Majorana case, not only is $B_z=\epsilon=0$, but the in-plane direction of the field (\emph{i.e.} the phase of $h_1$) is fixed, corresponding to the conservation of a dual fermion parity shared by the Majoranas near the center junction. This leads to a precise probability of $\mathcal{P}_{00}=1/2$ \emph{independent of dynamics} in the case of true Majoranas. In the spin-1/2 picture, the field in the case of true Majorana modes is confined to (say) the x-direction, while in the case of `accidental' degeneracy $\epsilon=0$ the field is merely confined to the equatorial plane. In either case, if the spin is initially in the direction of the field and the field is quickly turned off, the spin remains in the plane no matter what path the field took in turning off. Likewise in either case if the field is varied infinitely slowly so that the spin remains locked to the field direction, the spin remains in the plane as the field turns off. In the intermediate case, however, the field that is confined only to the plane may deflect the spin out of the plane as it varies, leading to a $\mathcal{P}_{00}\neq1/2$, which is impossible if the field is only allowed to vary along a line. This analogy informs our discussion above on the importance of the `slowness' of the cutting protocol.  In this analogy, quasiparticle poisoning (or temperature effects) correspond to spin decoherence, which shortens the spin polarization vector within the Bloch sphere and ultimately leads to a $P_{00}=1/2$ result, also independent of any dynamics. These decoherence effects therefore set the ultimate lower bound on the precision with which the Majorana splitting may be measured.

\section{Braiding}\label{Sec: Braiding}

    \begin{figure}
    \includegraphics[trim=4cm 4cm 4cm 4cm, clip, width=\columnwidth]{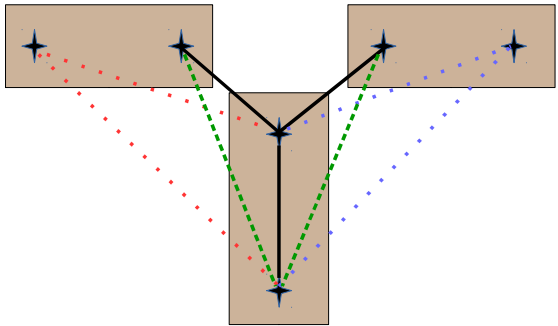}
        \caption{(Color Online) Diagram of the geometry assumed in our braiding analysis. We consider a three island geometry. The qubit is stored in the parity of superconducting islands 1 (left) and 3 (right), whose total parity is taken to be even at the beginning of the experiment. The middle island (2), is interposed between islands 1 and 3 so that no direct hopping from island 1 to island 3 is allowed, simplifying the analysis. In the topological case, each island would contain two Majorana zero modes, and only the nearest to the island junction would couple and decouple from one another during the braid process. The allowed couplings in the topological case are represented by solid black lines. We may represent the non-topological case in the same Majorana basis, but more couplings are now allowed between the Majorana modes. Case 1 allows all the couplings shown in solid, dashed and dotted lines. Case 2 separates out one Majorana mode on an outer island, disallowing either the red (left) or blue (right) set of dotted lines as couplings by setting $h_1=h_1'$ or $h_2=h_2'$ in Eq.~(\ref{eq:H}) respectively. Case 3 allows none of the dotted line couplings ($h_1=h_1'$, $h_2=h_2'$), and tends toward the topological case as the dashed green couplings are turned off ($\theta=\pm \pi/2$ in Eq.~(\ref{eq:H})).}\label{Fig:Aasen-geometry}
    \end{figure}
    We have thus far seen that the `fusion rule' based experiments described above, while an attractive stepping stone to braiding from an instrumentation standpoint, may not qualitatively identify topological systems more than the transport experiments that have already been conducted. (Certainly, fusion does not provide a sufficient condition for the existence of  non-Abelian Majorana modes since there are explicit situations, as discussed above, where purely non-topological systems with multiple low-lying fermionic modes would produce similar fusion signals.) Rather, we look to the braid properties of the Majorana system in order to demonstrate unequivocal topological behavior. As predicted in several places\cite{Nayak08,Alicea12a,Leijnse12,Beenakker2013a,Stanescu13,DasSarma15,Elliott15} and reviewed in Aasen et al.,\cite{Aasen16} a system of four Majorana zero modes, two each on left and right `islands', is expected to have the following behavior upon exchange of one of the Majoranas from the left island with one from the right: Before any exchanges, the fermion parity is set to be even on both islands, so that the probability $\mathcal{P}_{0}$ of finding both parities even is $\mathcal{P}_{0}=1$. After the first exchange, $\mathcal{P}_{0}=1/2$; after the second, $\mathcal{P}_{0}=0$; the third, $\mathcal{P}_{0}=1/2$; the fourth, $\mathcal{P}_{0}=1$ again and the sequence starts over. This may be summed up as
    \begin{equation}\label{Ptop}
      \mathcal{P}_{00}=\frac12+\frac12\cos{\frac{n\pi}{2}}
    \end{equation}
    where $n$ is the number of braid operations, all taken to wind in the same direction.
    \subsection{Single exchange}
        In order to conduct a braid operation with Majorana zero modes bound to semiconductor nanowires, one generically needs to leave a strictly one-dimensional setting\cite{Alicea11,Sau11b,Clarke11b}. Here, we follow the geometry of Aasen et al.,\cite{Aasen16} shown in Fig.~\ref{Fig:Aasen-geometry}, which allows some simplifications in the Hamiltonian we examine, although we expect our results to be broadly applicable. Out of the six Majorana bound states shown in Fig.~\ref{Fig:Aasen-geometry}, only the three at the ends of the islands toward the middle of the system participate in the braid process through direct tunneling in the ideal (topological) case, while the bottom Majorana bound state participates through the charging energy of the central island (island 2). In a more general (non-topological) case, couplings are allowed between any two Majorana modes, although we shall assume that the islands storing the qubit information at the beginning of the braid process have fermionic zero modes.

        In this section we consider in detail the Hamiltonian of this 3-island geometry as it is tuned through the braid process. Each of the islands will be assumed to have a single low-lying fermion state, and the total parity of the three islands will be assumed to be conserved over the duration of the experiment.  The most general Hamiltonian of this type, restricted to the even parity sector in which our computation takes place, is given by
        \begin{equation}
            H=\left(\begin{array}{c} \ket{000}\\\ket{011}\\\ket{110}\\\ket{101}\end{array}\right)^T
                \left(\begin{array}{cccc}
                    A & h_2 & h_1 & h_3\\
                    h_2^* & B & h_3' & h_1'\\
                    h_1^* & h_3'^* & C & h_2'\\
                    h_3^* & h_1'^* & h_2'^* & D\\
            \end{array}\right)\left(\begin{array}{c} \bra{000}\\\bra{011}\\\bra{110}\\\bra{101}\end{array}\right).
        \end{equation}
        We shall alternately write our Hamiltonian and unitary transformations in bra/ket notation or in matrix notation. Matrices in this section should be understood to be written in the above basis. The three entries in each bra or ket represent the fermion parity of islands 1-3 respectively.

        We may refine this Hamiltonian using the following assumptions:

        \textbf{Assumption 1}: Parity cannot hop directly from island 1 to island 3 (so $h_3=h_3'=0$).

        \textbf{Assumption 2}: The cross-capacitance of the three islands is small, so that we may assign independent energies $\epsilon_i$ to the occupation of the fermion states on islands $i=1,~2,~3$.

        \textbf{Assumption 3}: In the absence of tunnel coupling, islands 1 and 3 each have a fermionic zero mode (so $\epsilon_1=\epsilon_3=0$).

        Assumption 1 is justified in the braiding arrangement of Aasen et al, in which island 2, used for braiding, is interposed between islands 1 and 3. We may therefore take the direct coupling of islands 1 and 3 to be small. This assumption is specific to the geometry of Aasen et al., but direct coupling between islands 1 and 3 is not necessary to braiding, so we exclude it. Assumptions 2-3 are experimentally motivated by the idea that these wires have presumably already demonstrated zero-bias conductance peaks in transport experiments\cite{Mourik12,Das12,Deng12,Rokhinson12,Finck12,Churchill13,Chang15,Zhang16,Albrecht16} and potentially demonstrated near-zero-energy modes in fusion-rule experiments as well (as in Sec.~\ref{Sec: Fusion}), so that both the bare excitation energy and the capacitive energy may be assumed to be small.

        The most general Hamiltonian under these assumptions is
        \begin{equation}\label{eq:H}
            H=\left(\begin{array}{cccc}
                -\epsilon_2 & h_2 e^{i\theta} & h_1  & 0\\
                h_2 e^{-i\theta} & \epsilon_2 & 0 & h_1' \\
                h_1  & 0 & \epsilon_2 & h_2' e^{-i\theta}\\
                0 & h_1'  & h_2' e^{i\theta} & -\epsilon_2\\
            \end{array}\right).
        \end{equation} where we have performed a gauge transformation to make the $h_i$ and $h_i'$ entries real and positive. The matrix therefore contains only a single phase parameter $\theta$.

        To describe the braiding process, we further assume that one may independently decouple the first island ($h_1=h_1'=0$) or the third island ($h_2=h_2'=0$), or cause a degeneracy in the second island ($\epsilon_2=0$).

        The braiding process begins with all the couplings off ($h_i=h_i'=0$), and island 2 given a charging energy so that $\epsilon_2\neq=0$. We then proceed in the following steps:

        \textbf{Step 1} Decrease $\epsilon_2$ to 0, while increasing the tunnel coupling between islands 1 and 2 so that \mbox{$h_1,~h_1'\rightarrow h_{1max}\neq0$}.

        \textbf{Step 2} Decrease $h_1,~h_1'$ to 0, while increasing the tunnel coupling between islands 2 and 3 so that \mbox{$h_2,~h_2'\rightarrow h_{2max}\neq0$}.

        \textbf{Step 3} Decrease $h_2,~h_2'$ to 0, while increasing the charging energy on island 2 so that $\epsilon$ returns to near its original value.

        We assume that these steps are performed adiabatically, so that each eigenstate at the beginning of a step is mapped to an eigenstate at the end of that step in such a way that the ordering of the energies is maintained.

        Step 1 begins with two degenerate states ($\ket{000}$ and $\ket{101}$, but there is no ambiguity in the mapping because the Step 1 Hamiltonian does not couple these two sectors.
        We may thus infer that the adiabatic performance of Step 1 results in the unitary operation:
        \begin{eqnarray}
          U_1&=&\frac{e^{i\xi_1}}{\sqrt{2}}(\ket{000}-\ket{110})\bra{000}\nonumber\\
             &+&\frac{e^{i\xi_2}}{\sqrt{2}}(\ket{011}+\ket{101})\bra{011}\nonumber\\
             &+&\frac{e^{i\xi_3}}{\sqrt{2}}(\ket{000}+\ket{110})\bra{110}\nonumber\\
             &+&\frac{e^{i\xi_4}}{\sqrt{2}}(\ket{011}-\ket{101})\bra{101},\nonumber\\
        \end{eqnarray}
        where the $\xi_i$ are accumulated phases (a combination of dynamic and Berry phases) to be determined later if necessary. Note that the ground state degeneracy is maintained during Step 1 only if $h_1=h_1'$

        Likewise, performing Step 3 adiabatically results in:
        \begin{eqnarray}
          U_3&=&\frac{e^{i\psi_1}}{\sqrt{2}}\ket{000}(\bra{000}-e^{i\theta}\bra{110})\nonumber\\
             &+&\frac{e^{i\psi_2}}{\sqrt{2}}\ket{011}(\bra{000}-e^{i\theta}\bra{110})\nonumber\\
             &+&\frac{e^{i\psi_3}}{\sqrt{2}}\ket{110}(e^{i\theta}\bra{110}+\bra{101})\nonumber\\
             &+&\frac{e^{i\psi_4}}{\sqrt{2}}\ket{101}(e^{i\theta}\bra{110}-\bra{101}),\nonumber\\
        \end{eqnarray}
        where again the $\psi_i$ are accumulated phases to be determined later if necessary. Note that the ground state degeneracy is maintained during Step 3 only if $h_2=h_2'$

%         The Step 2 Hamiltonian has eigenvalues $\pm\lambda_\pm$, where
%         \begin{eqnarray}
%           \lambda_\pm^2&=&\frac{h_1^2+h_2^2+h_1'^2+h_2'^2}{2}\nonumber\\
%           &\pm& \frac12 \sqrt{(h_1^2-h_1'^2)^2+(h_2^2-h_2'^2)^2+2(h_1^2+h_1'^2)(h_2^2+h_2'^2)+8h_1 h_2 h_1' h_2' \cos 2\theta}
%         \end{eqnarray}
        The analysis of Step 2 is somewhat more involved. During this step, degeneracy is maintained between the lowest two energy states only if $h_1=h_1'$, $h_2=h_2'$ and $\theta=\pm \pi/2$. These are exactly the conditions for the topological system. Any non-topological system will accumulate a dynamic phase due to the splitting. Furthermore, if $\theta=0,\pi$, the first and second excited energy states meet during the evolution, which leads to diabatic errors in the evolution--the system will not return to the ground state space after the braid operation. This type of error can be somewhat mitigated by measuring all three islands at the end of the braid test, rather than just those containing the qubit.\cite{Knapp16}

        If adiabatic evolution is maintained, however, we note three distinct cases for a general non-topological system: $h_{1,2}$ are each distinct from $h_{1,2}'$, one of them is distinct, or both are equal. These three cases lead to quite different unitary transformations. Note that these requirements are not sharp, and that the crossover between the regimes below is governed by the rate at which the braid process is performed. For instance, in what follows, $h_1=h_1'$ may be taken in practice to mean $\abs{h_1-h_1'}\ll \hbar/\tau$, where $\tau$ is the timescale over which the braid is performed, while $h_1\neq h_1'$ indicates $\abs{h_1-h_1'}\gg \hbar/\tau$. We do not consider the case $\abs{h_1-h_1'}\sim \hbar/\tau$, where a consideration of the full non-adiabatic dynamics of the system is required and the results are expected to interpolate between those below. We also ignore constraints from any quasiparticle poisoning in our braiding consideration assuming such poisoning time scales to be very long compared with all braiding operation time scales (See Sec.~\ref{Sec: Poisoning} for the effects of quasiparticle poisoning)

        \subsubsection{Case 1: Generic couplings ($h_1\neq h_1'$ and $h_2\neq h_2'$)}

            We first consider the case in which all of the couplings shown in Fig.~\ref{Fig:Aasen-geometry} are allowed. Dotted-line (non-topological) couplings between islands are assumed to be zero whenever the solid-line couplings between those islands are off. The generic case in which all the couplings shown in Fig.~\ref{Fig:Aasen-geometry} are allowed corresponds to $h_1\neq h_1'$ and $h_2\neq h_2'$ in Eq.~(\ref{eq:H}). In this case there is no degeneracy at any time during Step 2. We may then proceed with the same analysis as worked for Steps 1 and 3 above, as each eigenstate is mapped unambiguously by adiabatic evolution. There are, however, two possibilities as to the order of the energies at each end of Step 2, leading to different transformations. Either $\sgn(h_1-h_1')=\sgn(h_2-h_2')$ or $\sgn(h_1-h_1')=-\sgn(h_2-h_2')$.
            In the first case,
            \begin{eqnarray}
                U_2&=&\frac{e^{i\zeta_1}}{2}(\ket{000}-e^{i\alpha}\ket{110})(\bra{000}-\bra{110})\nonumber\\
                   &+&\frac{e^{i\zeta_2}}{2}(\ket{000}-e^{i\alpha}\ket{110})(\bra{011}+\bra{101})\nonumber\\
                   &+&\frac{e^{i\zeta_3}}{2}(e^{i\alpha}\ket{110}+\ket{101})(\bra{000}+\bra{110})\nonumber\\
                   &+&\frac{e^{i\zeta_4}}{2}(e^{-i\alpha}\ket{110}-\ket{101})(\bra{000}-\bra{101}),\nonumber\\
            \end{eqnarray}
            leading to an overall unitary $U=U_3U_2U_1$ given by
            \begin{eqnarray}
                U=\left(\begin{array}{cccc}
                    e^{i\alpha_1} & 0 & 0  & 0\\
                    0 & e^{i\alpha_2} & 0 & 0 \\
                    0  & 0 & e^{i\alpha_3} & 0\\
                    0 & 0  & 0 & e^{i\alpha_4}\\
                \end{array}\right)\nonumber\\
            \end{eqnarray}
            that simply adds a phase $\alpha_i=\psi_i+\zeta_i+\xi_i$ to each of the initial eigenstates. We label this possibility Case~1a in later sections.

            The other possibility, that $\sgn(h_1-h_1')=-\sgn(h_2-h_2')$, ultimately leads to a deterministic switching between the ground states
            \begin{eqnarray}
                U=P_f\left(\begin{array}{cccc}
                    0 & 0 & 0  & 1\\
                    0 & 0 & 1 & 0 \\
                    0  & 1 & 0 & 0\\
                    1 & 0  & 0 & 0\\
                \end{array}\right)P_i\nonumber\\
            \end{eqnarray}
            at the completion of each nominal braid, where $P_f$ and $P_i$ are diagonal unitary matrices collecting the dynamic and Berry phases. We label this possibility Case~1b in later sections.
        \subsubsection{Case 2: One isolated Majorana mode ($h_1=h_1'$ and $h_2\neq h_2'$)}

            We now move on to the case in which either the left most or right-most Majorana mode in Fig.~\ref{Fig:Aasen-geometry} is isolated from the rest of the system. In this case, a degeneracy exists within the even-parity subsector at one end of Step 2.\footnote{The existence of at least one isolated Majorana mode implies that the even and odd parity sectors are degenerate, but not that a degeneracy exists \emph{within} a parity sector.} We take $h_1=h_1'$, with $h_2\neq h_2'$, corresponding to an isolated Majorana on the left island and the absence of the dotted (red) couplings to the far left side of that island in Fig.~\ref{Fig:Aasen-geometry} (the analysis of $h_1\neq h_1'$, $h_2= h_2'$, an isolated Majorana on the right island, is similar and omitted here). Physically, this situation might arise if one wire is fully topological, so that the situation corresponds to isolated or exponentially-protected MZMs, while the other two wires simply have near-zero-energy modes but are in the topologically trivial phase (\emph{e.g.} two nearby Majorana bound states with nearly orthogonal wave functions). Now there is a degeneracy at the beginning of Step 2, so that it is no longer immediately clear how adiabatic evolution will map the eigenstates. However, except in the topological case (where degeneracy persists throughout Steps 1, 2 and 3), the degeneracy is immediately lifted upon adding a small $h_2\sim h_2'\sim\hbar/\tau\ll h_{1max}$. Furthermore, barring the special case $\theta=0,\pi$ or a fully topological system, there are no gap closures during the evolution. We can therefore find the initial eigenvectors using degenerate perturbation theory for small $h_2 \sim h_2'$ and proceed as we did in Case 1.

            The energy eigenstates near the beginning of Step 2 and their corresponding eigenvalues are given by
            \begin{eqnarray}
                \frac12\left(\begin{array}{c} e^{i\phi}\\ -1 \\ -e^{i\phi}\\1\end{array}\right),~~-h_1-\eta_2;&\quad&
                \frac12\left(\begin{array}{c} e^{i\phi}\\ 1 \\ -e^{i\phi}\\-1\end{array}\right),~~-h_1+\eta_2;\nonumber\\
                \frac12\left(\begin{array}{c} e^{i\phi}\\ -1 \\ e^{i\phi}\\-1\end{array}\right),~~h_1-\eta_2;&\quad&
                \frac12\left(\begin{array}{c} e^{i\phi}\\ 1 \\ e^{i\phi}\\1\end{array}\right),~~h_1+\eta_2.\nonumber\\
            \end{eqnarray}
            Here \mbox{$\eta_2=\sqrt{h_2^2+h_2'^2+2h_2h_2'\cos 2\theta}+\mathcal{O}(h_2^2)$} and \mbox{$\phi=\arg(h_2e^{i\theta}+h_2'e^{-i\theta})$}.

             Adiabatic evolution maps these states to the eigenstates of the Hamiltonian with $h_1=\epsilon=0$. Taking $h_2>h_2'$, we have
             \begin{eqnarray}
                U_2&=&\frac{e^{i\zeta_1}}{2\sqrt{2}}\left(\begin{array}{c} 1\\ -e^{-i\theta} \\
                       0\\0\end{array}\right)\left(\begin{array}{cccc} e^{-i\phi}& -1 & -e^{-i\phi}&1\end{array}\right)\nonumber\\
                   &+&\frac{e^{i\zeta_2}}{2\sqrt{2}}\left(\begin{array}{c} 1\\ e^{-i\theta} \\
                       0\\0\end{array}\right) \left(\begin{array}{cccc} e^{-i\phi}& 1 & e^{-i\phi}&1\end{array}\right)\nonumber\\
                   &+&\frac{e^{i\zeta_3}}{2\sqrt{2}}\left(\begin{array}{c} 0\\ 0 \\ e^{-i\theta}
                       \\1\end{array}\right)\left(\begin{array}{cccc} e^{-i\phi}& -1 & e^{-i\phi}&-1\end{array}\right)\nonumber\\
                   &+&\frac{e^{i\zeta_4}}{2\sqrt{2}}\left(\begin{array}{c} 0\\ 0 \\ e^{-i\theta}
                       \\-1\end{array}\right)\left(\begin{array}{cccc} e^{-i\phi}& 1 & -e^{-i\phi}&-1\end{array}\right)\nonumber\\,
             \end{eqnarray}
             so that the total unitary enacted by the nominal braid is given by
             \begin{eqnarray}
                U=\frac{1}{\sqrt{2}}P_f
                \left(\begin{array}{cccc}
                    e^{-i\phi}& 0 & 0           &-1\\
                    0 &         1 & e^{-i\phi}  & 0\\
                    0 &        -1 & e^{-i\phi}  & 0\\
                    e^{-i\phi}& 0 & 0           & 1\\
                \end{array}\right)\nonumber\\
             \end{eqnarray}
             where $P_f$ is a diagonal unitary matrix collecting the dynamic and Berry phases. Note that no dynamic phase is accumulated during Step 1,\footnote{The Berry phase accumulated there is also zero in the gauge used here.} so that $P_i=I$.
        \subsubsection{Case 3: Two isolated Majorana modes ($h_1=h_1'$ and $h_2= h_2'$)}
            We now move on to the case in which both of the wires storing the initial state of the qubit are fully topological. In Fig.~\ref{Fig:Aasen-geometry} the dotted (red and blue) couplings to the far ends of the outer islands are disallowed. The consequence for the Hamiltonian stems from the fact that each of these wires has a Majorana zero mode that is far away from the junction, and so does not enter into the Hamiltonian for the braid process. With the appropriate choice of gauge, this means that we may set $h_1=h_1'$ and $h_2= h_2'$. Now there is a degeneracy at both ends of Step 2. This two-ended degeneracy allows the possibility of interference effects between different paths in the Hilbert space, which is crucial to the topological case. The present case differs from the topological case in that $\theta\neq\pm\pi/2$. This represents the non-topological nature of the center island, and results in energy splitting during Step 2. Following a procedure similar to that detailed for Case 2 and finding the appropriate eigenstates at each end of Step 2, we find a total unitary given by
            \begin{eqnarray}
                U=
                \left(\begin{array}{cccc}
                    \cos\frac{\chi}{2}& 0 & 0           & i\sin\frac{\chi}{2}\\
                    0 &     \cos\frac{\chi'}{2}     & i\sin\frac{\chi'}{2} & 0\\
                    0 &        i\sin\frac{\chi'}{2} & \cos\frac{\chi'}{2}  & 0\\
                    i\sin\frac{\chi}{2}& 0 & 0           & \cos\frac{\chi}{2}\\
                \end{array}\right),\nonumber\\
             \end{eqnarray}
             where $\chi,~\chi'$ include Berry and dynamic phases. As the dynamic phases tend to zero, $\chi,~\chi'\rightarrow\pm\pi/2$, the topological result.  That is, the Berry phase associated with the braid process always gives a $\pm\pi/2$ contribution to $\chi,~\chi'$, regardless of the dynamic phase. Assuming that no other states are excited, this result may also be achieved by performing the nominal braid quickly relative to the induced gap. Note that no dynamic phase\footnote{or Berry phase, in this gauge} is accrued during Steps 1 and 3, where the degeneracy of the qubit states is maintained.

    \begin{table*}[htb]
    {\setlength{\topsep}{-\parskip}
    \setlength{\partopsep}{0pt}
    \begin{tabularx}{\textwidth}{|l||X|X|X|X|X|X|}
    \hline
     \underline{Case} & Isolated Majorana modes & z(0) & z(1)  & z(2) & z(3) & z(4) \\
    \hline
     1a & 0 & 1 & 1 & 1 & 1 & 1\\
    \hline
     1b & 0 & 1 & -1 & 1 & -1 & 1\\
    \hline
     2 & 1 & 1 &  0 &$-e^{-\frac{\sigma_2^2}{2}}\cos\bar{\chi}_2$ & $e^{-\sigma_2^2}\sin^2\bar{\chi}_2$ & $e^{-\sigma_2^2}\cos^2\bar{\chi}_2+$ $e^{-\frac{3\sigma_2^2}{2}}\cos\bar{\chi}_2\sin^2\bar{\chi}_2$\\
    \hline
     3 & 2 & 1 & $e^{-\frac{\sigma_3^2}{2}}\cos \bar{\chi}_3$ & $e^{-\sigma_3^2}\cos 2\bar{\chi}_3$ & $e^{-\frac{3\sigma_3^2}{2}}\cos 3\bar{\chi}_3$& $e^{-2\sigma_3^2}\cos 4\bar{\chi}_3$\\
    \hline
     4 & 2 & 1 & 0 & -1 & 0 & 1\\
    \hline
    \end{tabularx}}
    \caption{Qubit polarization $z(n)=2\mathcal{P}_{00}-1$ after $n$ repeated braid attempts. Cases 1-3 are detailed in the main text and Fig.~\ref{Fig:Aasen-geometry}. Case 4 is the topological case, listed here for reference. The second column lists the number of Majorana modes that remain isolated, not participating in the braid process.  The mean $\bar{\chi}_i$ and variance $\sigma_i^2$ of the dynamic phases are system dependent. Note that $\bar{\chi}_2=0$, $\sigma_2=0$ makes the Case~2 results equal the topological ones, and $\bar{\chi}_3=\pm\frac{\pi}{2}$, $\sigma_3=0$ gives the series of results associated with a topological system in Case~3. Neither of these cases are topological. However, these are not equivalent situations. Case~3 is a `near topological' system, and tends to Case~4 as the bottom Majorana on the center island of Fig.~\ref{Fig:Aasen-geometry} becomes better isolated during Step 2 and therefore $\bar{\chi_3}\rightarrow\pm\frac{\pi}{2}$ with zero variance. The topological seeming result in Case~2 is somewhat coincidental and should change drastically with a small change in experimental parameters.  }
    \label{Table1}
    \end{table*}
    \subsection{Qubit polarization after multiple braids}
        Thus far, we have discussed the possible unitary transformations resulting from carrying out a braid process (\emph{i.e.} the set of operations that enacts a braid on a topological system) when the experimental system is not necessarily topological. In this, we are able to see the entirety of the expected unitary. An experiment verifying the braid process is generally much more limited, measuring only the probability of, \emph{e.g.}, returning to the state $\ket{000}$ after completing the braid process some number of times. We label this probability $\mathcal{P}_0(n)$, where $n$ is the number of nominal braids and the subscript indicates that an even parity is measured on both wires. That is, $\mathcal{P}_0(n)$ is the conditional probability of measuring $0$ on the left wire given that the overall parity is measured to be even.\footnote{This definition becomes important in the presence of quasiparticle poisoning, as our definition requires that $\mathcal{P}_{0}(n)+\mathcal{P}_{1}(n)=1$ (i.e. non-parity conserving results are thrown out).}

        To find $\mathcal{P}_0(n)$, we first reduce the unitary braid transformations of Sec~\ref{Sec: Braiding} to the parity conserving sector, giving the following four options:
        \begin{eqnarray}
            \mathrm{Case~1a}&&\nonumber\\
                U&=&\left(
                \begin{array}{cc}
                    e^{-i\chi_{1a}/2}  & 0\\
                    0 &  e^{i\chi_{1a}/2}\\
                \end{array}\right)\nonumber\\
            \mathrm{Case~1b}&&\nonumber\\
                U&=&\left(
                \begin{array}{cc}
                    0 & -e^{-i\chi_{1b}/2} \\
                    e^{i\chi_{1b}/2} &  0\\
                \end{array}\right)\nonumber\\
            \mathrm{Case~2}&&\nonumber\\
                U&=&\frac{1}{\sqrt{2}}\left(
                \begin{array}{cc}
                    e^{-i\frac{\chi_2}{2}}& -e^{-i(\frac{\chi_2}{2}-\phi)}\\
                    e^{i(\frac{\chi_2}{2}-\phi)}&  e^{i\frac{\chi_2}{2}}\\
                \end{array}\right)\nonumber\\
            \mathrm{Case~3}&&\nonumber\\
                U&=&\left(
                \begin{array}{cc}
                    \cos\frac{\chi_3}{2}&  i\sin\frac{\chi_3}{2}\\
                    i\sin\frac{\chi_3}{2}&  \cos\frac{\chi_3}{2}\\
                \end{array}\right).\nonumber\\
        \end{eqnarray}
        In each case we have removed all of the overall phases in favor of a single remaining dynamic phase variable $\chi_i$. We assume the experimental procedure is the same for each attempted braid, so that for each attempt $\chi$ is a random variable with mean $\bar{\chi}_i$ and variance $\sigma^2_i$. \footnote{This is an appropriate assumption if, e.g. $\abs{h_2'-h_2}\tau/\hbar\gg 1$, which is both our Case 2 condition and an indication that the dynamic phase winds many times around $2\pi$ over the duration of the braiding procedure.} After each nominal braid, the density matrix transforms as
        \begin{equation}
            \rho(n)=\int\frac{\mathrm{d}\chi}{\sqrt{2\pi}\sigma_i} U(\chi)\rho(n-1) U^\dagger(\chi)e^{-\frac{(\chi-\chi_0)^2}{2\sigma_i^2}},
        \end{equation}
        where $n$ is the total number of nominal braid operations (of the same chirality) performed.

        We define the qubit polarization
        \begin{equation}
            z(n)=\bra{000}\rho(n)\ket{000}-\bra{101}\rho(n)\ket{101})
        \end{equation}
        so that
        \begin{equation}
            \mathcal{P}_0(n)=\frac12+\frac12 z(n),
        \end{equation}
        where $\mathcal{P}_0(n)$ is the probability of finding the system in the state $\ket{000}$ after $n$ nominal braids.

        Assuming we begin with the system in the ground state with even parity on both islands storing the qubit ($\ket{000}$), we find
        \begin{eqnarray}
            \mathrm{Case~1a}&&\nonumber\\
                z(n)&=&1\\
            \mathrm{Case~1b}&&\nonumber\\
                z(n)&=&(-1)^{n}\\
            \mathrm{Case~2}&&\nonumber\\
                z(0)&=&1\nonumber\\
                z(1)&=&0\nonumber\\
                z(2)&=&-e^{-\frac{\sigma_2^2}{2}}\cos\bar{\chi}_2\nonumber\\
                z(3)&=&e^{-\sigma_2^2}\sin^2\bar{\chi}_2\nonumber\\
                z(4)&=&e^{-\sigma_2^2}\cos^2\bar{\chi}_2+e^{-3\frac{\sigma_2^2}{2}}\cos\bar{\chi}_2\sin^2\bar{\chi}_2\nonumber\\
            \mathrm{Case~3}&&\nonumber\\
                z(n)&=&e^{-\frac{n\sigma_3^2}{2}}\cos n\bar{\chi}_3.\nonumber\\
        \end{eqnarray}
        For Case~2, we have listed only the polarization values for the first four braids of the same chirality. A solution for arbitrary $n$ is possible, but cumbersome, and we do not show it here. The Case~3 results reduce to those (Eq.~\ref{Ptop}) for a fully topological system when $\bar{\chi}_3=\pm\pi/2$, $\sigma_3^2=0$. Note, however, the possibility of an accidental similarity between Case 2 results and those of true braiding, should $\bar{\chi}_2\approx 0$. The case 2 results after a single braid are also noteworthy: A probability of $\mathcal{P}_0=.5$ is obtained independent of the dynamic phase distribution in this (non-topological) case! This is another example of the unreliability of 50-50 type signals in attempting to distinguish topological phenomena.

        \begin{figure}
          \includegraphics[trim=4cm 4cm 4cm 4cm, clip, width=\columnwidth]{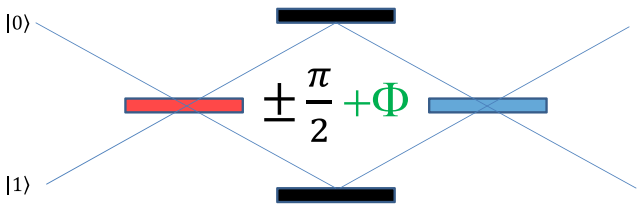}
          \caption{(Color Online) A linear optics device analogous in operation to the braid process. When both the left-and right-hand optical elements are 50-50 beam splitters, and the accumulated phase in the resulting interferometer loop is $\pm\pi/2$, an incoming beam is transformed by the topological braid matrix (as in our case 3). Non-topological couplings between the islands in Fig.~\ref{Fig:Aasen-geometry} lead to imperfections in the corresponding optical elements, so that, e.g. generic couplings (red in Fig.~\ref{Fig:Aasen-geometry}) leading to $h_1\neq h_1'$ cause the left hand beam splitter to  be either replaced by a mirror or removed entirely (in either case, a random dynamic phase is added). Likewise the (blue in Fig.~\ref{Fig:Aasen-geometry}) non-topological couplings to the right hand island replace the right hand beam splitter with a mirror or remove it. Finally, the inability to isolate the bottom Majorana on the center island in Fig.~\ref{Fig:Aasen-geometry} leads to a non-topological value of the phase $\Phi$ accumulated in the interferometer.
          All of our braid results are reproduced by this device.}
          \label{Fig:interferometer}
        \end{figure}
        The three cases may be compactly summarized by an analogy with a linear optics device, as shown in Fig~\ref{Fig:interferometer}. In particular, this gives insight into the oddly exact value $z(1)=0$ found in Case~2. This case is analogous to an interferometer in which one of the two 50-50 beam splitters has been replaced by a mirror, so that the beam is always split evenly into the two outputs regardless of the accumulated phase.  We have further summarized our results for the first four braids for easy reference in Table~\ref{Table1}.

        We have not discussed or included any effect of extrinsic decoherence in the above analysis.  In practice, of course, all dynamic processes involving braiding and fusion will be limited by an effective decoherence time which puts a sharp lower (upper) limit on how slow ($\tau$) the braiding can be in realistic situations.  In particular, the whole braiding measurement must be carried out over a time scale much shorter than the typical decoherence time (e.g. the poisoning time associated with the entrance of stray quasiparticles in the system from the environment). Heuristically, however, one might expect the effect of such poisoning to be similar to that of variance in the dynamic phase in Cases~2 and~3 --the signal is degraded toward an unpolarized ($z=0$) state. We expect such depolarizing noise to be present in all Cases, further complicating the distinction between the topological and non-topological results.

\section{Quasiparticle poisoning}\label{Sec: Poisoning}
    Since topological qubits are intrinsically robust, the dominant `decoherence' mechanisms are extrinsic, with quasiparticle poisoning (i.e. introduction of random stray non-thermal quasiparticles into the superconductor) being the most important one.  Such quasiparticles violate parity conservation, destroying the topological qubit.
    While great strides have been made in limiting the problem of quasiparticle poisoning in Majorana nanowires experimentally, \cite{vanWoerkom15,Higginbotham15} the presence of stray fermions within the superconducting system will have a deleterious effect on braiding experiments. We approach this problem with the simplifying assumption that stray quasiparticles will populate all fermionic modes within the system equally, regardless of energy, since they are intrinsically nonthermal in nature (all thermal quasiparticles are exponentially supppressed at low temperatures due to the superconducting gap). In particular, if the system begins in an even state of total fermion parity, then \emph{no information about the braid process remains if the total measured fermion parity is odd}. Indeed, as the extra fermion is equally likely to have entered the left or right wires, whether the prior state was $\ket{00}$ or $\ket{11}$ is completely unknown. However, in the sector with total even fermion parity, information degrades more slowly. There, the net effect of quasiparticle poisoning (as with many decohering processes) is a tendency to drive $\mathcal{P}_0\rightarrow 1/2$. This form of quasiparticle poisoning leads to the master equation for the density matrix $\rho^e$ of the even subspace
    \begin{equation}
        \dot{\rho}^e=-i[H,\rho^e]-\frac{\kappa_p}{2}(\rho^e-\frac{\mathrm{Tr}\rho^o}{N_e}I)
    \end{equation}
    where $\rho^o$ is the density matrix of the odd subspace, $I$ is the identity matrix, and $N_e$ is the size of the even parity subspace (in our simple model for the braiding experiment, $N_e=4$, while $N_e=2$ for the simplest model of the fusion experiment). $\kappa_p$ is the poisoning rate. Note that only the trace of $\rho^o$ enters the equation for $\rho^e$, representing the fact that all information (other than the total density) is lost in the poisoning process.

    This master equation is actually separable: noting that $\mathrm{Tr}\rho^o=1-\mathrm{Tr}\rho^e$ and setting
    \begin{equation}
        \rho^e=\frac{\mathrm{Tr}\rho^e}{N_e}I+e^{-\frac{\kappa_p t}{2}}(\bar{\rho}-\frac{1}{N_e}I),
    \end{equation}
    where
    \begin{equation}
        \mathrm{Tr}\dot{\rho}^e=-\kappa_p(\mathrm{Tr}\rho^e-\frac12)
    \end{equation}
    we find that $\bar{\rho}$ is a unit trace matrix obeying
    \begin{equation}
        \dot{\bar{\rho}}=-i[H,\bar{\rho}].
    \end{equation}
    We may therefore treat the quasiparticle poisoning separately from the Hamiltonian evolution, as if it all occurred (e.g.) at the end of the braid process.

    Given that no information about the braid process is transferred into the odd parity sector, we assume all islands involved in the experiment are measured (including the central island in the braiding experiment), and that results showing an odd total parity are thrown out. We must therefore normalize the probability to only these outcomes.

    Beginning with a state such that at $t=0$, $\mathrm{Tr}\rho^e=1$, we find for an $n$-braid process that takes time $t$
    \begin{eqnarray}
        P^{\mathrm{poisoned}}_{00}(n)&=&\frac{\rho^e_{000}}{\rho^e_{101}+\rho^e_{000}}\nonumber\\
                                %&=&\frac{\frac{1+e^{-\kappa t}}{8}+e^{-\frac{\kappa t}{2}}(\bar{\rho}_{000}-\frac{1}{4})}{\frac{1+e^{-\kappa t}}{4}+e^{-\frac{\kappa t}{2}}(\bar{\rho}_{000}+\bar{\rho}_{101}-\frac{1}{2})}\nonumber\\
                                %&=&\frac{\frac{1+e^{-\kappa t}}{8}+e^{-\frac{\kappa t}{2}}(\frac14+z(n)/2)}{\frac{1+e^{-\kappa t}}{4}+e^{-\frac{\kappa t}{2}}(\frac12)}\nonumber\\
                                &=&\frac12+\frac{z(n)}{\cosh(\frac{\kappa_p t}{2})+1}
    \end{eqnarray}
    where $\rho^e_{s}=\bra{s}\rho^e\ket{s}$. Note that we have taken \mbox{$\bar{\rho}_{000}+\bar{\rho}_{101}=1$} and $\bar{\rho}_{000}=(1+z(n))/2$ at the end of the braid process, thus assuming no diabatic errors in the Hamiltonian evolution.
    Obviously, the outcome depends crucially on the poisoning rate $\kappa_p$, which must be a phenomenological input into the theory. However, as might be expected, the poisoned system has a tendency to revert to $P^{\mathrm{poisoned}}_{00}=1/2$ over time, independent of the braiding operation that is performed. This underlines the importance of control experiments in which no braid is expected to be performed, especially in any experiment for which the expected topological result is $P_{00}=1/2$. We emphasize that for finite quasiparticle poisoning rate, both braiding and fusion invariably produce a 1/2-probability if the experiment is conducted over a long time period. This is independent of the topological nature of the system!  Thus, only experiments done over a time scale fast compared with quasiparticle poisoning would be relevant for the observation of Majorana zero modes in fusion or non-Abelian statistics in braiding.
\section{Summary and Outlook}\label{Sec: Outlook}
    We have critically considered recently proposed fusion and braiding experiments in semiconductor nanowires, carefully distinguishing features arising from trivial non-topological effects (which could nevertheless be quite subtle) from the important ones arising from topological effects of non-Abelian Majorana statistics, finding that there are situations where the same experimental signature could arise equally well from both topological and non-topological origins, making the situation less clear-cut or definitive in the presence of realistic effects which are likely to be present experimentally.  We then discussed various specific techniques capable of discerning non-Abelian signatures in braiding (or fusion) from trivial non-topological effects.  In general, our conclusion is that fusion, by itself, can only satisfy necessary conditions supporting the existence of non-Abelian excitations, probing essentially the same information as the existing Majorana signatures manifested in the zero bias conductance peak in nanowire tunneling transport measurements (although fusion could provide compelling support for Majorana modes if successful in nanowires already manifesting zero bias tunneling conductance peaks, paricularly because it has the potential to provide better energy resolution than transport experiments).  We have carried out a detailed analysis of various proposed nanowire braiding experiments clarifying in depth the non-Abelian signatures as distinguished from accidental non-topological effects, thus providing a clear guideline for future experiments on how to discern non-Abelian braiding statistics from the measurements. Below we provide a more technical summary of our findings, emphasizing our assumptions and approximations, and discuss their implications briefly.

    Our results are obtained within a simple model that assumes only a few low-energy modes are present in the system. In the case of fusion, we find that even if there is only a single mode per wire segment, it is possible to reproduce the topological results of fusion with non-topological system whose on-site energy is pinned near zero.  Additional modes do not provide further distinction and in fact tend to converge to the topological result in the many-mode limit. In the case of a braiding experiment, we find that the model with a single complex fermionic mode per island is sufficient to provide a distinction between the topological case and the case of non-topological fermions with on-site energy fine-tuned to zero.
    Our results suggest an `operational' definition of a non-Abelian Majorana zero mode at a given endpoint as one whose partner is completely decoupled from any operator located at that endpoint (couplings all much less than the inverse timescale of the experiment). That is, any coupling to that end of the nanowire by fermionic modes outside the wire `sees' only one Majorana mode. This is a stronger condition than the decoupling of Majorana modes within the wire, as evidenced by the variety of braiding results possible even when such decoupling is present (see Sec.~\ref{Sec: Braiding}). This condition also leads to the `true' Majorana result for the fusion experiment of Sec.~\ref{Sec: Fusion} (\emph{i.e.} that the $P_{00}=1/2$ result occurs independent of the dynamics), although as described in that section this result my be difficult to distinguish from those of more generic systems. In order to conclusively establish the existence of such anyonic Majorana zero mode excitations, it is important that the experimental signatures satisfy both necessary and sufficient conditions for non-Abelian braiding statistics and not just the necessary conditions (since the necessary conditions may also arise from non-topological effects in various circumstances as discussed in our work).

    It is important to note in either the fusion or braiding context that the measurements of the system are necessarily destructive. After each measurement, the system is reset. In particular, this means that a measurement of $z(3)$ (\emph{i.e.} measuring the qubit polarization after three braids) is three times as costly in time as a measurement of $z(1)$ (the polarization after just one braid). This may indeed be far too costly, increasing the possibility of decoherence effects such as quasiparticle poisoning substantially.  However, one braid is not the best option even though it can obviously be carried out faster than a multiple braid experiment. The probability $\mathcal{P}_{0}(n)$ is probed through a series of Bernoulli trials, ultimately resulting in a binomial distribution for the qubit state (0 or 1) with mean \mbox{$\mathcal{P}_{1}(n)=1-\mathcal{P}_{0}(n)$} and variance $\mathcal{P}_{0}(n)\mathcal{P}_{1}(n)/N_\mathrm{tr}$, where $N_\mathrm{tr}$ is the number of trials. In order to distinguish one of the non-topological cases from the topological one, enough statistics must be accumulated to separate the resulting probability distributions. Judging by Table~\ref{Table1}, this may be particularly difficult after just one braid. Case~2 is indistinguishable from the topological case after a single braid, while Case~3 may give $\mathcal{P}_{0}(1)=1/2$ for non-topological reasons if the variance is large. (In fact, in the presence of depolarizing noise such as quasiparticle poisoning, all braid results have a tendency to revert to $\mathcal{P}_0=1/2$). The result after a double braid is more robust because the topological result lies at one extreme of the range of values for $z$. Dephasing, noise, quasiparticle poisoning and non-topological terms can only move the polarization \emph{away} from the topological result for a double braid.

    In general we find that experiments with a predicted unpolarized outcome of $z=0$ (including fusion experiments) do not robustly distinguish topological and non-topological systems, at least not without careful manipulation of the involved system parameters, such as wire length or experimental timescale. This is not to say that such experiments have no value. Indeed the timescale of the cut in a fusion experiment may be used to put a bound on the energy splitting of Majorana modes in the associated double wire system. In particular, such fusion measurements, if successful, provide evidence in support of Majorana zero modes at the same level of satisfying a necessary condition as do the zero bias conductance peak observation. As fusion is an independent measurement not connected with tunneling transport, this would further strengthen the case for non-Abelian statistics (while not clinching it conclusively). Likewise, the braiding experiment should be carefully optimized to include data from both the first and second braids (and beyond, if possible) in distinguishing between the cases described here.

    Finally, we comment on four concrete aspects of the experimental nanowires which are likely to be important physical mechanisms in determining success or failure of fusion/braiding experiments.  First, the current experimental nanowires manifest zero bias peaks whose broadening or energy width is typically comparable to the topological gap.  If the zero bias peak width arises entirely from splitting between the Majorana bound states, then there is the serious problem of fusion/braiding experiment being unable to satisfy the key necessary condition of being much faster (slower) than the Majorana splitting (topological gap), as these energy scales are not well separated in current transport experiments\cite{Mourik12,Das12,Deng12,Rokhinson12,Finck12,Churchill13,Chang15,Zhang16,Albrecht16}.  Second, it has recently been argued\cite{DasSarma16, Liu16} that intrinsic dissipative broadening may play a role in Majorana nanowires.  In the presence of such dissipation it is of course necessary for fusion/braiding to occur much faster than the dissipation energy scale, which may be difficult to satisfy until dissipative broadening is suppressed in the experimental systems compared with the current estimates.  Third, all measurements must occur on a time scale much faster than any quasiparticle poisoning time scale, which necessitate careful engineering to avoid the nonthermal poisoning endemic in mesoscopic superconducting structures. Finally, much like quasiparticle poisoning, finite temperature will generically degrade the visibility of fusion  or braiding results, drawing the measured probabilities toward the maximumally random result of $P=1/2$. It is interesting to note that such a result, in the absence of an appropriate control experiment,\cite{Aasen16} would lead to false positives (suggesting a topological result in a non-topological system) in the fusion experiment but false negatives (a non-topological result in a topological system) in a double-braid experiment.

    Our work shows that a fusion experiment, when interpreted cautiously, could lead to useful information regarding the existence of non-Abelian Majorana zero modes, but, by itself, cannot conclusively establish non-Abelian statistics because the possibility of trivial non-topological effects can never be completely ruled out.  Braiding measurements can of course distinguish the topological phase from the non-topological ones, but sufficient statistics involving many measurements with different numbers of braids may be necessary in order to make a compelling case for the existence of non-Abelian statistics.
\section{Acknowledgements}
    The authors would like to thank Jason Alicea for several thought provoking discussions.
    This work is supported by Microsoft and by the Laboratory for Physical Sciences (LPS-MPO-CMTC)
\bibliography{topo-phases-1-24-17}
\end{document}